\documentclass[aps,prl,twocolumn,superscriptaddress,preprintnumbers,amsmath,amssymb,floatfix,showpacs,showkeys,nofootinbib,reprint]{revtex4-1}

\usepackage{graphicx,fink}
\usepackage{epsfig}
\usepackage{dcolumn}
\usepackage{bm}
\usepackage{longtable}
\usepackage{color}
\usepackage{float}
\usepackage{comment}
\usepackage{amsmath}
\usepackage{amsfonts}
\usepackage{ulem}
\usepackage{xspace}
\usepackage{ulem}

\newcommand{\ChiEFT}{\ensuremath{\chi}EFT\xspace}

\newcommand{\nuc}[2]{\ensuremath{^{#1}}#2}
\newcommand{\geant} {{\bf {G}\texttt{\scriptsize{EANT}}4}}

\begin{document}

\preprint{MAX-lab $^{2}$H$(\gamma,\gamma)$ Summary Article for Physical Review Letters}

\title{New Measurement of Compton Scattering from the Deuteron and an Improved Extraction of the Neutron Electromagnetic Polarizabilities}

\author{\mbox{L.~S.~Myers}} \altaffiliation{Present address: Thomas Jefferson
  National Accelerator Facility, Newport News, VA 23606, USA}
\affiliation{Department of Physics, University of Illinois at
  Urbana-Champaign, Urbana, IL 61801, USA}
\author{\mbox{J.~R.~M.~Annand}} \affiliation{School of Physics and Astronomy,
  University of Glasgow, Glasgow G12 8QQ, Scotland UK}
\author{\mbox{J.~Brudvik}} \affiliation{MAX IV Laboratory, Lund University,
  SE-221 00 Lund, Sweden}
\author{\mbox{G.~Feldman}} 
\affiliation{Institute for Nuclear Studies, Department of Physics, The George
  Washington University, Washington, DC 20052, USA}
\author{\mbox{K.~G.~Fissum}} \altaffiliation{Corresponding author;
  \texttt{kevin.fissum@nuclear.lu.se}} \affiliation{Department of Physics,
  Lund University, SE-221 00 Lund, Sweden} 
\author{\mbox{H.~W.~Grie\3hammer}}
\affiliation{Institute for Nuclear Studies, Department of Physics, The George
  Washington University, Washington, DC 20052, USA} \author{\mbox{K.~Hansen}}
\affiliation{MAX IV Laboratory, Lund University, SE-221 00 Lund, Sweden}
\author{\mbox{S.~S.~Henshaw}} 
\altaffiliation{Present address: National Security Technologies, Andrews AFB, MD 20762, USA} 
\affiliation{Department
  of Physics, Duke University, Durham, NC 27708, USA}
\author{\mbox{L.~Isaksson}} \affiliation{MAX IV Laboratory, Lund University,
  SE-221 00 Lund, Sweden} \author{\mbox{R.~Jebali}} \affiliation{School of
  Physics and Astronomy, University of Glasgow, Glasgow G12 8QQ, Scotland UK}
\author{\mbox{M.~A.~Kovash}} \affiliation{Department of Physics and Astronomy,
  University of Kentucky, Lexington, KY 40506, USA}
\author{\mbox{M.~Lundin}} \affiliation{MAX IV Laboratory, Lund University,
  SE-221 00 Lund, Sweden}
\author{\mbox{J.~A.~McGovern}} \affiliation{School of Physics and Astronomy,
  The University of Manchester, Manchester M13 9PL, UK}
\author{\mbox{D.~G.~Middleton}} 
\affiliation{Kepler Centre for Astro and Particle Physics, Physikalisches
  Institut, Universit{\"a}t T{\"u}bingen, D-72076 T{\"u}bingen, Germany}
\author{\mbox{A.~M.~Nathan}} \affiliation{Department of Physics, University of
  Illinois at Urbana-Champaign, Urbana, IL 61801, USA}
\author{\mbox{D.~R.~Phillips}} \affiliation{Department of Physics and
  Astronomy and Institute of Nuclear and Particle Physics, Ohio University,
  Athens, Ohio 45701, USA} \author{\mbox{B.~Schr\"oder}}
\affiliation{MAX IV Laboratory, Lund University, SE-221 00 Lund, Sweden}
\affiliation{Department of Physics, Lund University, SE-221 00 Lund, Sweden}
\author{\mbox{S.~C.~Stave}} \altaffiliation{Present address: Pacific Northwest
  National Laboratory, Richland, WA 99352, USA} \affiliation{Department of
  Physics, Duke University, Durham, NC 27708, USA}

\collaboration{The COMPTON@MAX-lab Collaboration} \noaffiliation

\date{\today}

\begin{abstract}
The electromagnetic polarizabilities of the nucleon are fundamental properties that describe its response to external electric and magnetic fields. They can be extracted from Compton-scattering data---and have been, with good accuracy, in the case of the proton. In contradistinction, information for the neutron requires the use of Compton scattering from nuclear targets.
Here we report a new measurement of elastic photon scattering from deuterium using quasimonoenergetic tagged photons at the MAX IV Laboratory in Lund, Sweden.
These first new data in more than a decade effectively double the 
world dataset. Their energy range overlaps with previous experiments and
extends it by $20$~MeV to higher energies. An analysis using Chiral
Effective Field Theory with dynamical $\Delta(1232)$ degrees of freedom
shows the data are consistent with and within the world dataset. After demonstrating that the fit is
consistent with the Baldin sum rule, extracting values for the isoscalar
nucleon polarizabilities and combining them with a recent result for the
proton, we obtain the neutron
polarizabilities as 
$\alpha_n=[11.55\pm1.25(\text{stat}) \pm 0.2(\text{BSR}) \pm
0.8(\text{th})]\times10^{-4}\,\text{fm}^3$ and $\beta_n=[3.65 \mp
1.25(\text{stat}) \pm 0.2(\text{BSR}) \mp
0.8(\text{th})]\times10^{-4}\,\text{fm}^3$,
with $\chi^2=45.2$ for $44$ degrees of freedom. 


\end{abstract}


\maketitle

%%%%%%%%%%%%%%%%%%%%%%%%%%%%%%%%%%%%%%%%%%%%%%%%%%%%%%%%%%%%%%%%%%%%
% begin body
%%%%%%%%%%%%%%%%%%%%%%%%%%%%%%%%%%%%%%%%%%%%%%%%%%%%%%%%%%%%%%%%%%%%

The electric and magnetic dipole polarizabilities $\alpha$ and $\beta$ of the 
proton and neutron have recently drawn much attention; see 
e.g.~\cite{griesshammer2012} for a review and~\cite{letter} for an open letter.
They encode the response of the nucleon to applied electric or magnetic fields,
and summarize information on the 
entire spectrum of nucleonic excitation, offering a stringent test of our understanding 
of Quantum Chromodynamics (QCD). Full lattice QCD results with realistic uncertainties 
are  anticipated in the near future~\cite{lujan2014,*detmold2012}.
Besides being fundamental nucleon properties, $\alpha$ and $\beta$ are relevant for
theoretical studies of the Lamb shift of muonic hydrogen and of the
proton-neutron mass difference, and dominate the uncertainties of both
\cite{Pachucki:1999,Carlson:2011dz,Birse:2012eb,WalkerLoud:2012bg}.

The majority of nucleon-polarizability measurements have used nuclear Compton scattering. 
This paper reports new results for the deuteron, where elastic Compton scattering provides 
access to the isoscalar average of proton and neutron polarizabilities, $\alpha_s$ and $\beta_s$.
A review of the three previous measurements of this process~\cite{lucas1994,hornidge2000,lundin2003} 
can be found in Ref.~\cite{griesshammer2012}. Prior to our measurement, the
database consisted of 29 points between $49$ 
and $95$~MeV. The best extant Chiral Effective Field Theory (\ChiEFT) 
calculation~\cite{griesshammer2012} 
fits these data with a $\chi^2$ per degree of freedom ($\chi^2/{\rm d.o.f.}$) of $0.97$ 
and gives [in units of $10^{-4}$~${\rm fm}^3$, used throughout]
\begin{equation}
  \label{eq:AlphaBetaS1}
  \alpha_s - \beta_s = 7.3 \pm 1.8(\text{stat}) \pm 0.8(\text{th}).
\end{equation}

Here the Baldin Sum Rule (BSR)~\cite{baldin1960}, a variant of 
the optical theorem, was used to constrain the fit. Evaluating the sum rule using proton photoabsorption 
cross-section data gives $\alpha_p + \beta_p=13.8 \pm 0.4$~\cite{deleon2001}. 
For the neutron, the requisite cross sections are found from empirical 
partial-wave amplitudes for pion photoproduction. 
We take $\alpha_n + \beta_n=15.2 \pm 0.4$~\cite{levchuk2000} with 
the uncertainty highly correlated with that for the proton, and so
\begin{equation}
  \label{eq:BSRS}
 \alpha_s + \beta_s=14.5 \pm 0.4.
\end{equation}

This  world database of deuteron Compton scattering
is much smaller, is of poorer quality, and spans a
narrower energy range than that for the proton. A statistically consistent proton
Compton database up to 170~MeV contains 147 points~\cite{McGovern:2012ew},
with its most recent \ChiEFT fit of
\begin{equation}
  \label{eq:AlphaBetaP}
  \begin{split}
    &\alpha_p = 10.65 \pm 0.35(\text{stat}) \pm 0.2(\text{BSR}) \pm
    0.3(\text{th}) \\ 
    &\beta_p = 3.15 \mp 0.35(\text{stat}) \pm 0.2(\text{BSR}) \mp
    0.3(\text{th}),
  \end{split}
\end{equation}
using the proton BSR.
Combining (\ref{eq:AlphaBetaS1}), (\ref{eq:BSRS}), 
and (\ref{eq:AlphaBetaP}) yields
\begin{equation}
  \label{eq:AlphaBetaN}
  \begin{split}
    &\alpha_n = 11.1 \pm 1.8(\text{stat}) \pm 0.2(\text{BSR}) \pm
    0.8(\text{th}) \\ 
    &\beta_n = 4.1 \mp 1.8(\text{stat}) \pm 0.2(\text{BSR}) \mp 0.8(\text{th}).
  \end{split}
\end{equation}
These numbers are consistent with
an extraction of neutron polarizabilities from 7 data points
on \nuc{2}{H}($\gamma$,$\gamma^\prime$$n$)$p$, taken on the quasielastic ridge above 
$200$~MeV~\cite{Kossert:2002ws}.
Again using the neutron BSR constraint, this gives:
\begin{equation}
  \label{eq:AlphaBetaKossert}
  \begin{split}
    &\alpha_n = 12.5 \pm 1.8(\text{stat}){}^{+1.1}_{-0.6}(\text{sys})\pm
    1.1(\text{th}) \\ 
    &\beta_n = 2.7 \mp 1.8(\text{stat}){}_{-1.1}^{+0.6}(\text{sys})\mp
    1.1(\text{th}).
  \end{split}
\end{equation}
where the theory uncertainty may be underestimated~\cite{Lvov}.

As the statistical uncertainties dominate the overall uncertainty in Eq.~(\ref{eq:AlphaBetaS1}), 
there is a pressing need for more and better deuteron Compton data.
In this Letter, we report a new and comprehensive measurement of the
differential cross section for elastic Compton scattering from deuterium performed 
at the MAX IV Laboratory~\cite{m4,eriksson2014} in Lund, Sweden. This measurement nearly doubles the 
number of data points in the world database and enables us to extract $\alpha_n$ and $\beta_n$ 
with statistical uncertainties which are substantially reduced compared to 
those of Eq.~\eqref{eq:AlphaBetaN}.

At the Tagged-Photon Facility~\cite{tpf,adler2012} at the MAX IV Laboratory, 
we used a 15~nA, 45\% duty factor pulse-stretched electron
beam~\cite{lindgren2002} with energies of 144~MeV and 165~MeV 
to produce quasimonoenergetic photons in the energy range 65 -- 115 MeV via 
the
bremsstrahlung-tagging technique~\cite{adler1990,adler1997}. This range covers
most previous experiments and extends the range towards higher energies by
$20$~MeV. The post-bremsstrahlung electrons were momentum analyzed in a
magnetic spectrometer and detected in a 62-channel scintillator hodoscope
\cite{vogt1993} located along the focal plane (FP). The resulting
tagged-photon beam had an energy resolution of $\sim$500~keV per channel and a mean flux of
$\sim$2$\times 10^6$~MeV$^{-1}$s$^{-1}$. The collimated photon beam was
incident on a scattering chamber containing liquid deuterium in a
cylindrical cell (length 170~mm and diameter 68~mm) made from 120~$\mu$m
Kapton foil. The thickness of the target was (8.10$\pm$0.20) $\times$
10$^{23}$ nuclei/cm$^2$.  The average loss of incident-beam photons due to
absorption in the target was approximately 2\%. The tagging
efficiency~\cite{adler1997} was the ratio of the number of tagged photons
which survived the collimation and struck the target to the number of
post-bremsstrahlung electrons which were registered by the associated FP
channel. It was monitored on a daily basis using very low-intensity beam and a
Pb-glass detector with 100\% efficiency for the photons of interest. The
tagging efficiency was determined to be (44 $\pm$ 1$_{\rm sys}$)\%.

Three large-volume, segmented NaI(Tl)
detectors~\cite{miller1988,wissmann1994,myers2010} were used to detect the
Compton-scattered photons. The detectors were each composed of a single, large
NaI(Tl) core crystal surrounded by optically-isolated, annular NaI(Tl)
segments. The detectors had an energy resolution of better than 2\% FWHM at
energies near 100 MeV which enabled the separation of elastically
scattered events from events due to deuterium breakup. The signals from the
NaI(Tl) detectors were passed to charge-integrating analog-to-digital
converters (QDCs) and time-to-digital converters (TDCs) and the data were
recorded on an event-by-event basis. The QDCs allowed reconstruction of the
scattered-photon energies, while the TDCs provided the time correlation between 
the NaI(Tl) detectors and the FP hodoscope. The data-acquisition system was
triggered by an event in any one of the NaI(Tl) detectors which gated the QDCs
and started the TDCs used for the timing measurements. The TDC
stop signals came from the FP detectors. The energy calibration of each
detector was determined by placing it directly into a low-intensity photon
beam and observing the response as a function of tagged-photon energy.
The energy calibration was confirmed to $\sim$1\% with the 
131.4~MeV photon from the capture of $\pi^{-}$ on deuterium and 
reconstruction of the $\pi^{0}$ mass in back-to-back kinematics 
as defined by opposing NaI detectors.

\begin{figure}[h]
  \begin{center}
    \includegraphics[width=0.5\textwidth]{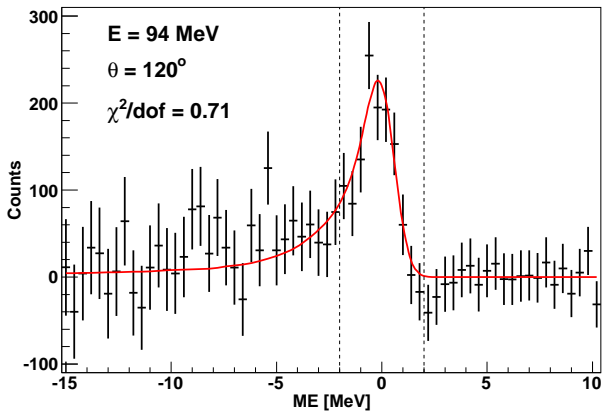}
\caption{
(Color online) A typical accidental-corrected scattered-photon spectrum
together with a simulation (red) of the response function of the detector.
The vertical, dashed lines indicate the $-$2.0 $<$ ME $<$
2.0 MeV yield-integration region. 
\label{figure:figure_01_scattering_peak}} 
\end{center}
\end{figure}

In scattering configuration, the detectors were located at laboratory angles
of 60$^\circ$, 120$^\circ$, and 150$^\circ$ and at corresponding distances of
34.3, 41.8, and 91.5~cm from the target. Data were collected over
two separate 4-week run periods.
Gain instabilities in the NaI(Tl) detectors were corrected using the location
of the QDC peak for selected cosmic rays (those that pass through the diameter
of the core crystal) on a run-by-run basis. Missing
energy (ME) was defined as the difference between the energy registered in the
NaI(Tl) detector and the tagged-photon energy corrected for the Compton
scattering energy shift.  
\geant~\cite{agostinelli2003} simulations 
of the detector lineshapes 
were empirically broadened to 
match the measured in-beam detector responses.
Scattering-configuration \geant\ simulations of the
{\it in situ} detector responses, acceptances, and efficiencies were based on the
broadened in-beam results and were used to determine the total yield in the
elastic-scattering peak. The sum of the resulting \geant\ lineshape and an
accidental background was fit to the data. The \geant\ simulation was also used
to correct for the detection efficiency of the NaI(Tl) detector over a region
of interest (ROI) of $-$2.0 $<$ ME $<$ 2.0 MeV, the ME integration region used
to determined the yield. The ROI was dictated by the resolution of the
detectors and the 2.2~MeV threshold for the breakup of deuterium. 
The fitting procedure was repeated with a second lineshape 
that simulated photons from \nuc{2}{H}($\gamma$,$\gamma^\prime$)$n$$p$. The 
contribution of these photons to the extracted yield in the ROI was consistent 
with zero. Additionally, extraction of the cross section for a slightly wider or
narrower ROI produced results in agreement with those presented here.
Effects associated with the finite size of the experimental apparatus
as well as a
correction for scattered photons absorbed by the target ($\sim$1\%) were also
simulated. A typical accidental-corrected scattered-photon spectrum, the
\geant\ simulation of the response function of the detector, and the
integration region are shown in Fig.~\ref{figure:figure_01_scattering_peak}.
The simulation is clearly in good agreement with the data over the ROI 
($\chi^2$/d.o.f. = 0.71).

The integrated scattered-photon yield was normalized to the number of photons
incident on the target and corrected for rate-dependent
factors~\cite{myers2013} which were due to the physical overlap of the FP
counters. This procedure has been systematically verified in measurements of
photon scattering from carbon~\cite{myers2014,preston2013}. The number of
photons incident on the target was determined from the number of
post-bremsstrahlung electrons detected in each FP channel, the tagging
efficiency, and the rate-dependent correction factors.

The correlated systematic uncertainties for the experiment arise from the
tagging efficiency (1\%), target thickness (3\%), target-cell contributions
(3\%), and rate-dependent effects (4\%). Angle-dependent effects are the 
product of the solid angle and detection efficiency (3\% at $150^\circ$ and
$4.2\%$ otherwise). Point-to-point
systematic uncertainties are dominated by the yield extraction 
($\sim$5\%).
A detailed discussion of the uncertainties, as well as the  
cross sections, can be found in Ref.~\cite{myers2014a}.
For the first data-production run, correlated systematic uncertainties add in
quadrature to $5.2\%$; and
for the second run, they add to 4.7\%.
A table of cross sections is provided in Ref.~\cite{supmat}.

The extraction of $\alpha_s$ and $\beta_s$ from  deuteron Compton scattering cross  
sections is not straightforward. Even for the case of the proton, most of the 
world data is beyond the energy at which a low-energy expansion of the cross 
section is valid. At energies above 100~MeV, the energy-dependent effects of the 
pion cloud and of the $\Delta(1232)$ excitation become important. Furthermore, the response 
of the deuteron to electric and magnetic fields is not just that of the constituent 
proton and neutron; for instance, pion-exchange currents contribute a substantial fraction 
of the deuteron Compton cross section at these energies~\cite{Beane99}.
\ChiEFT is ideally suited to account consistently for both these aspects since it 
encodes the correct symmetries and degrees of freedom of QCD in model-independent 
Compton-scattering amplitudes with systematically improvable theoretical uncertainties.  
It predicts the full energy dependence of the single-nucleon scattering response 
(including spin-dependent amplitudes).  For the deuteron, it consistently accounts for 
nuclear binding and obtains the correct Thomson limit by including NN 
rescattering~\cite{Hildebrandt05}.
  
Here we summarize the ingredients of our recent \ChiEFT analysis at order
$e^2\delta^3$ (next-to-leading order in $\alpha$ and $\beta$), as detailed in
Sect.~5.3 of Ref.~\cite{griesshammer2012}. The degrees of freedom are:  point nucleons
with anomalous magnetic moments; a dynamical $\Delta(1232)$, treated
nonrelativistically and without width; and the chiral pion clouds of both the proton
and $\Delta(1232)$ at their respective leading orders. Two short-distance contributions to 
$\alpha_s$ and $\beta_s$ are the only free parameters in our theory, since the 
$\gamma N\Delta$ $M1$ coupling is determined from proton Compton scattering.
We compute the photon-deuteron scattering amplitude to $O(e^2\delta^3)$, and so 
include all these one- and two-nucleon effects. The dependence of our results 
on the deuteron wave function and $NN$ potential is negligible.

We now fit $\alpha_s$ and $\beta_s$ using this theory. We use the same fit procedure
and parameters as in Ref.~\cite{griesshammer2012} and further details will appear in an upcoming 
publication~\cite{futuretheory}. Our fit adds point-to-point and angle-dependent 
systematic uncertainties in quadrature to the statistical uncertainty, and subsumes overall 
systematic uncertainties into a floating normalization (see Eq.~(4.19) of 
Ref.~\cite{griesshammer2012} and references therein). The deuteron Compton database 
of Ref.~\cite{griesshammer2012} is augmented by the two experimental runs reported here, which are 
treated as separate data sets with independent floating normalizations.
Treating them as one single data set does not significantly affect the results.
The theoretical uncertainty in the extracted polarizabilities from contributions 
beyond chiral order $e^2\delta^3$ has been assessed as $\pm0.8$ ~\cite{griesshammer2012}. 

Within the statistical uncertainties, consistent results are obtained whether we
analyze the new data alone, or in conjunction with the previous world data. Here
we present results only for the latter. 
In either case, the total $\chi^2$ receives an unacceptably 
large contribution from two points: $94.5$~MeV, $60^\circ$ and $112.1$~MeV, 
$120^\circ$. These individually contribute at least 8.4 and 10.8 to the total 
$\chi^2$, respectively (the exact contributions depend partially on fit details). 
The next largest contribution from a single datum is less than 4.6. Standard 
hypothesis-testing techniques thus require these two points to be excluded
if the data set is to be statistically consistent.
Fitting both the polarizabilities, we find
\begin{equation}
  \label{eq:AlphaBetaNewFree}
  \begin{split}
    &\alpha_s = 11.1 \pm 0.9(\text{stat})\pm0.8(\text{th})  \\ 
    &\beta_s = 3.3 \pm 0.6(\text{stat})\pm0.8(\text{th}) ,
  \end{split}
\end{equation}
with $\chi^2=49.2$ for $43$ degrees of freedom. 
This is in very close agreement with the isoscalar BSR~[Eq.~\eqref{eq:BSRS}], which
we can therefore use to reduce
the statistical uncertainties.
We then find
\begin{equation} 
  \label{eq:AlphaBetaSNew}
  \alpha_s-\beta_s=7.8 \pm 1.2(\text{stat}) \pm 0.8(\text{th}),
\end{equation}
which agrees well with Eq.~\eqref{eq:AlphaBetaS1}. This result leads to
\begin{equation} 
  \label{eq:AlphaBetaNewBaldin}
  \begin{split}
    &\alpha_s = 11.1 \pm 0.6(\text{stat}) \pm 0.2(\text{BSR}) \pm
    0.8(\text{th}) \\
    &\beta_s = 3.4 \mp 0.6(\text{stat}) \pm 0.2(\text{BSR}) \mp
    0.8(\text{th}).
  \end{split}
\end{equation}

\begin{figure}[!htb]
  \begin{center}
    \includegraphics[width=0.4\textwidth]{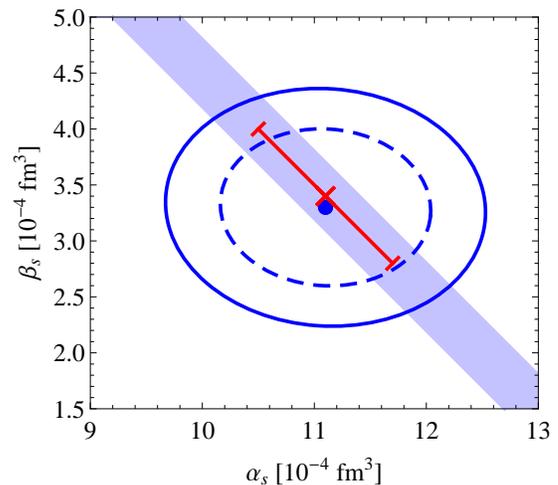}
    \caption{ (Color online) The $1\sigma$ ($\chi^2+2.3$, solid ellipse) and
      $\chi^2+1$ (dashed ellipse) regions of the free fit compared to the
      Baldin-constrained fit (line); grey band: BSR.}
\label{fig:chisquared}
\end{center}
\end{figure}

\begin{figure}[h]
  \begin{center}
    \includegraphics[width=0.48\textwidth,trim=25mm 50mm 25mm 22mm,clip]{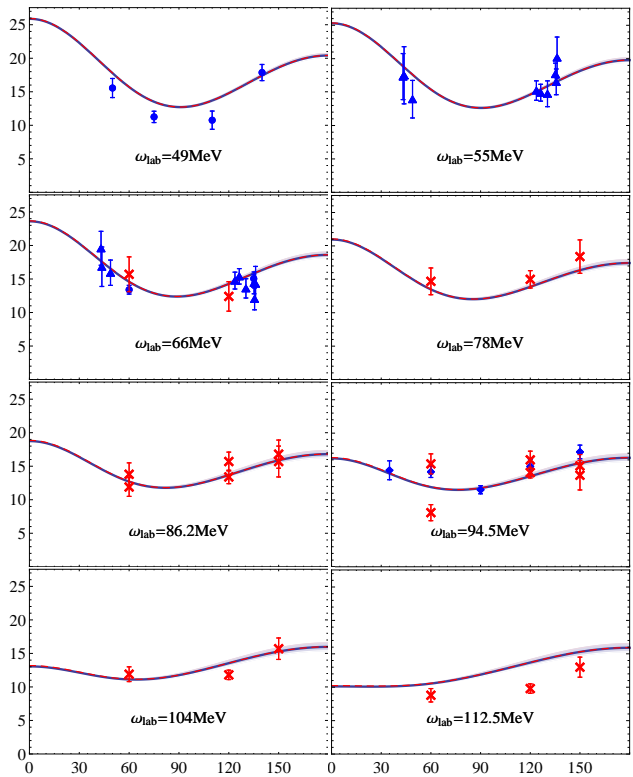}
\caption{
(Color online) Deuteron Compton scattering cross sections $[\mathrm{nb/sr}]$
in the lab frame as function of lab angle $[\mathrm{deg}]$. Data are within 2~MeV of the nominal energy.
Data and statistical uncertainties: this work $\textcolor{red}{\times}$; Ref.~\cite{lucas1994}
$\textcolor{blue}{\bullet}$; Ref.~\cite{hornidge2000}
$\textcolor{blue}{\blacklozenge}$; Ref.~\cite{lundin2003}
$\textcolor{blue}{\blacktriangle}$. Curves: results of the one/two parameter
fits (solid/dashed); band: statistical uncertainty
 of the one-parameter fit. Note that the two
points with $\Delta \chi^2 > 5$ are not included in the fit, and the normalization of each 
data set is floated within quoted uncertainties. See text for details.
\label{figure:figure_02_results}}
\end{center}
\end{figure}

The total $\chi^2$ is now $45.2$ for $44$ degrees of freedom. 
If we were to reinstate the two outliers, the central values would be 
11.5 and 3.0, with $\chi^2=70.2$.
We emphasize that the new data decrease the statistical uncertainty by $33\%$.
We observe that the overall normalization of each dataset floats by less than 5\%, indicating 
good absolute cross-section normalizations. The
$\chi^2$/d.o.f. of previous sets barely changes when the new data are added.
Further details will be given in Ref.~\cite{futuretheory}.
Figure~\ref{fig:chisquared} shows the $1\sigma$ ellipses of the free and
Baldin-constrained fits. Cross sections and fits are shown in Fig.~\ref{figure:figure_02_results}.

In order to extract neutron polarizabilities, we combine the isoscalar values with the 
proton polarizabilities from Eq.~\eqref{eq:AlphaBetaP}  to obtain
\begin{equation}
  \label{eq:AlphaBetaNnew}
  \begin{split}
    &\alpha_n = 11.55 \pm 1.25(\text{stat}) \pm 0.2(\text{BSR}) \pm
    0.8(\text{th}) \\ 
    &\beta_n = 3.65 \mp 1.25(\text{stat}) \pm 0.2(\text{BSR}) 
    \mp 0.8(\text{th}).
  \end{split}
\end{equation}
The shift in the central values from the previous extraction,
Eq.~\eqref{eq:AlphaBetaN}, is statistically insignificant, but our data shrink
the statistical uncertainty by $30\%$. The result is  in good agreement
with that from quasielastic scattering, Eq.~\eqref{eq:AlphaBetaKossert}. 

The BSR determinations of the proton and neutron $\alpha+\beta$ provide evidence 
for an isovector component of the sum of polarizabilities of around 10\% of the 
isoscalar part. In contrast, the present statistical and theoretical 
uncertainties on $\alpha-\beta$ 
are much too large to detect a significant proton-neutron difference. The
forthcoming extension of the higher-order \ChiEFT calculation on the 
proton~\cite{McGovern:2012ew} to the deuteron should reduce the theory  uncertainty 
substantially. However, the statistical uncertainty still dominates, so there is still a need 
for additional high-accuracy deuteron data.  Experiments are ongoing or 
planned at MAMI~\cite{annand2013}, the MAX IV Laboratory, and HI$\gamma$S~\cite{weller2009}. 
This continued effort to illuminate the structure of the nucleon through 
Compton-scattering measurements should soon directly confront lattice QCD 
extractions of $\alpha$ and $\beta$.

Here we have reported on an important step in this direction. Differential 
cross sections for elastic scattering from $^{2}$H have been measured using 
quasimonoenergetic tagged photons with energies in the range 65 -- 115 MeV at 
laboratory angles of 60$^\circ$, 120$^\circ$, and 150$^\circ$ at the Tagged-Photon 
Facility at the MAX IV Laboratory in Lund, Sweden. These data were used to extract
the isoscalar polarizabilities and reduced the statistical uncertainty on these
quantities by 33\%, thereby appreciably tightening the constraints on neutron 
structure from Compton scattering. 

%%%%%%%%%%% Acknowledgments

The authors acknowledge the support of the staff of the MAX IV Laboratory. We
also acknowledge the Data Management and Software Centre, a Danish
contribution to the European Spallation Source ESS AB, for providing access to
their computations cluster. 
We are grateful to the organisers and participants of the workshops 
\textsc{Compton Scattering off Protons and Light
     Nuclei: Pinning Down the Nucleon Polarizabilities}, ECT*, Trento (Italy, 2013), and
\textsc{Bound States and Resonances in Effective Field Theories and Lattice QCD Calculations}, 
Benasque (Spain, 2014).
The Lund group acknowledges the financial support
of the Swedish Research Council, the Knut and Alice Wallenberg Foundation, the
Crafoord Foundation, the Swedish Institute, the Wenner-Gren Foundation, and
the Royal Swedish Academy of Sciences. 
This material is based upon work supported by the National Science Foundation under Grant 0855569;
the U.S. Department of Energy, Office of Science, 
Office of Nuclear Physics under Award Numbers
DE-FG02-93ER40756,
DE-FG02-95ER40907,
 and
DE-FG02-06ER41422;
and the UK Science and Technology Facilities Council under Grants
ST/F012047/1 and ST/J000159/1.

\bibliography{myers_etal_v2}

\end{document}